# Extending the abilities of the Minkowski spacetime diagram

**Nilton Penha**
Departamento de Física, Universidade Federal de Minas Gerais, Brazil,
nilton.penha@gmail.com.

**Bernhard Rothenstein**
Politehnica University of Timisoara, Physics Department, Timisoara, Romania,
brothenstein@gmail.com.

**Doru Paunescu**
Politehnica University of Timisoara, Department of Mathematics, Timisoara, Romania

**Abstract**. A two-dimensional Minkowski spacetime diagram is neatly represented on a Euclidean ordinary plane. However the Euclidean lengths of the lines on the diagram do not correspond to the true values of physical quantities in spacetime, except for those referring to the stationary reference frame. In order to extend its abilities to other inertial reference frames, we derive a factor which, multiplied by the magnitude of the actually displayed values (on the diagram), leads to the corresponding true measured values by any other inertial observers. Doing so, the student can infer from the Euclidean diagram plot the expressions that account for Lorentz length contraction, time dilation and also Lorentz Transformations just by using regular trigonometry.

## 1. Introduction

Minkowski space-time diagrams are very helpful for understanding special relativity theory because they make transparent the relation between events and the parameters used to describe them by different reference frames.

An event is characterized by the space coordinates of the point where it happens and by its time coordinate *t* displayed by a clock located at the point considered. The set of all possible events is what is meant by spacetime. If Cartesian coordinates are used to specify the spatial localization of the event, the spacetime coordinates are *(x,y,z,t)* relative to a given reference frame. The values of the coordinates depend on the reference frame which is being used. Most of the textbooks use to represent the time coordinate component as *ct* where c is the speed of light in the vacuum, admitted as constant by the second postulate of special relativity; the first postulate is that "all physics laws are the same in all inertial frames".

In a two-dimensional Minkowski spacetime diagram, the spacetime is fully represented on a Euclidean plane (sheet of paper). This is possible because both the Euclidean points and the events can be labeled by pairs of real numbers. However there is an inherent limitation with the use of such diagrams. The Euclidean lengths of lines in the diagrams do not correspond to proper times or proper lengths in spacetime.

## 2. A thought experiment

Lets us consider a traditional thought experiment in which two observers $R_O$ and $R'_O$ move with constant relative velocity *V*. Let *K* be an inertial reference frame attached to $R_O$ and *K'* another inertial reference frame attached to $R'_O$. Then *K* and *K'* have a constant relative velocity *V*. In this case we don't need more than two spacetime coordinates *(x,ct)* and *(x',ct')*, the first pair associated with *K* and the second pair associated with *K'*. The *x-axis* and *x'-axis* are naturally taken along the line containing the two observers, the so called standard configuration.

In Figure 1 we show the scenario we are considering for discussion. The inertial reference frame *K* is arbitrarily considered stationary and reference frame *K'* is receding from *K* at velocity *V* along *x-axis*. All clocks in *K* are assumed synchronized according to Einstein criterion[1]. The two observers occupy, each one, the spatial origin of their reference frames and the time may be conveniently set to be zero when the spatial distance between them is also zero.

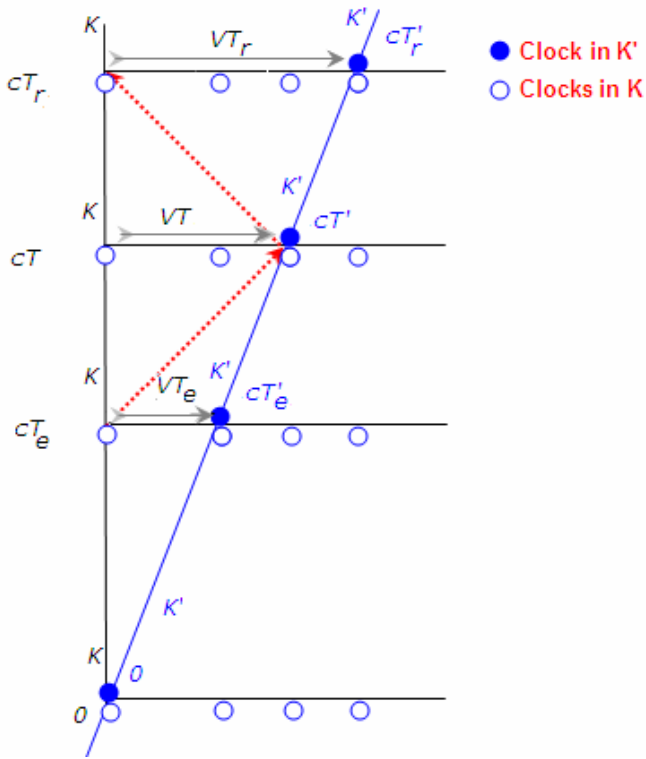

**Figure 1** – *K* is stationary and *K'* is receding from *K* at speed *V* along the horizontal direction.

Let us imagine that at time $cT_e$, a light pulse is emitted from a light source carried by $R_O$ in *K*. The light pulse reaches $R'_O$ at time $cT$, measured in *K*, and reflected back by a mirror at $R'_O$ place; the reflected light pulse reaches back *K* at time $cT_r$, measured in *K*. Here we list some of the events $E(x,ct)$ which happen in Figure 1 scenario:

*E(x=0,ct=0)* - both *K* and *K'* are coincident;

*E(x=0,ct=cT$_e$)* - the light pulse is emitted when all the clocks in *K* show time $ct= cT_e$;

*E(x=VTe,ct=cT$_e$)* - *K'* is at a spatial distance $x=VT_e$ from *K*;

*EM·(x=VT,ct= cT)* - the light pulse reaches the observer $R'_O$ (in *K'*) when all the clocks in *K* show time $ct=cT$ (this happens when spatial separation between them is $x=VT$);

*E(x=0,ct= cT$_r$)* - admitting that the light pulse is instantly reflected back, it reaches $R_O$ when all the clocks in *K* show time $ct=cTr$ (the echo time).

*E(x=VTr,ct= cT$_r$)* -  *K'* is at a spatial distance  $x=VT_r$  from *K*.

Let us find some relations between the above events. Since the light pulse which is emitted at time $cT_e$ reaches $K'$, the spatial separation between them is $x=VT$. So can write

$$VT = cT - cT_e. \qquad (1)$$

By the moment that the light pulse is back to $K$, the spatial separation between the two observers is $x=VT_r$. Then we have

$$VT = cT_r - cT. \qquad (2)$$

From the above relations (1) and (2) we obtain

$$cT = \frac{1}{2}(cT_r + cT_e), \qquad (3)$$

$$VT = \frac{1}{2}(cT_r - cT_e), \qquad (4)$$

$$cT = \frac{cT_e}{1-\beta}, \qquad (5)$$

$$cT = \frac{cT_r}{1+\beta}, \qquad (6)$$

$$\frac{cT_r}{cT_e} = \frac{1+\beta}{1-\beta}, \qquad (7)$$

where we use the notation $\beta=V/c$.

One interesting outcome of this experiment is that the observer $R_O$, which is at rest at the spatial origin in $K$, can measure the distance between him (her) and any object just by emitting a light pulse and having it back by reflection at the object in question; since the time at which the reflection occurs is given by (3), the distance he (she) is looking for is $x=(cT_r-cT_e)/2$. So just by recording the instant of emission $(cT_e)$ and the instant of the light pulse return $(cT_r)$ (echo), both spacetime coordinates $x$ and $cT$ of the reflection event are determined. This is so called radar technique[2] through which the spatial distances can be measured by clocks.

Every observer has his (her) own clock; it is very common to refer to observers' clock as his (her) wristwatch. So at the moment of the reflection, for example, while $R_O$'s wristwatch shows time $cT$, the one belonging to $R'_O$ shows time $cT'$. Both times are not equal; they differ by a constant factor which depends on the relative velocity.

## 3. Minkowski diagrams and Doppler factor

Let us represent the scenario of Figure 1 in terms of a diagram. The figure suggests drawing a vertical straight line to represent the succession of all events that happen at the same place, the position of $R_O$. We may call this line as the worldline of $R_O$, name it $WLR_O$. The worldline of $R'_O$, name it $WLR'_O$, is represented as a straight line which makes an angle $\theta$ with $WLR_O$ where

$$\tan\theta = \beta. \qquad (8)$$

The wordlines for the light pulses, $WL(+c)$ and $WL(-c)$, are drawn making an angle $+\pi/4$ and $-\pi/4$ with $WLR_O$, depending on if light recedes from $WLR_O$ to the right or to the left; $\tan(+\pi/4)$ and $\tan(-\pi/4)$ lead to $\beta=1$ (or $V=c$), and to $\beta=-1$ (or $V=-c$). A horizontal line represents a set of simultaneous events with respect to $WLR_O$.

This is an example of what is called Minkowski diagram. The worldline $WLR_O$ stands for *ct-axis* and the simultaneity line which intercepts $WLR_O$ at its origin stands for the *x-axis*. See Figure 2 for the diagram.

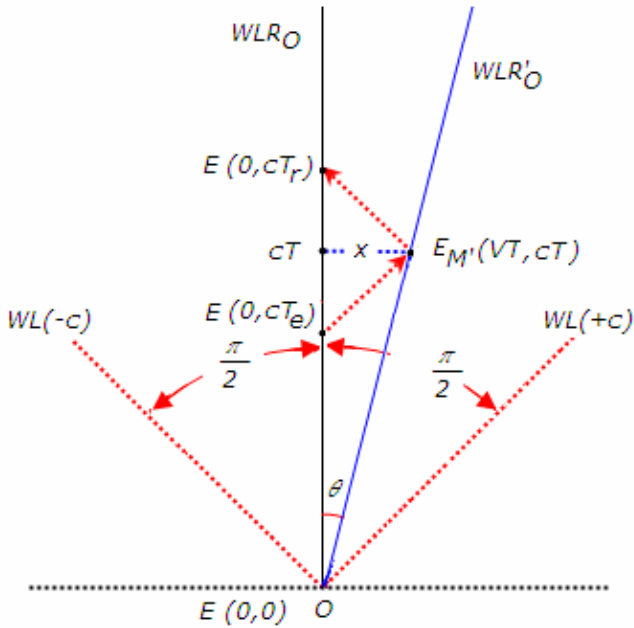 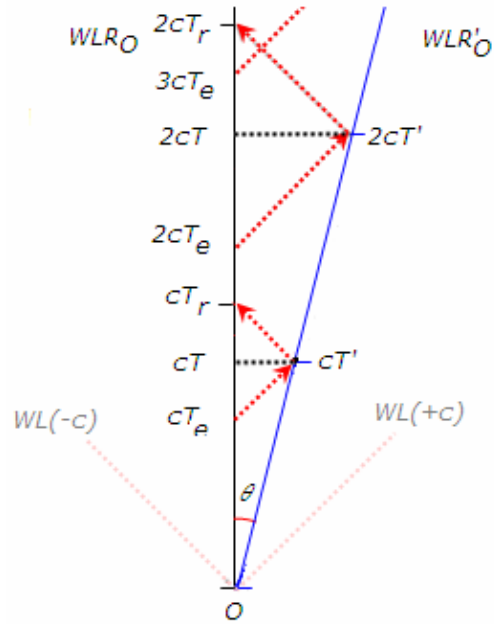

**Figure 2** - A spacetime diagram on the axes of which we measure the space coordinates *(x)* and the time coordinates *(ct)* of events. The worldline $WLR_O$ stands for *ct-axis* and the simultaneity line which intercepts $WLR_O$ at its origin stands for the *x-axis*.

**Figure 3** – Observer $R_O$, on $WLR_O$, emits a light pulse at every $cT_e$. The pulse is reflected back by observer $R'_O$, on $WLR'_O$, at every $cT$ ($cT'$) according to $R_O$ ($R'_O$). The echo reaches $R_O$, on $WLR_O$, at every $cT_r$.

Let us consider now that $R_O$, in our thought experiment, emits a light pulse at every $cT_e$ which is received back at every $cT_r$ after been reflected at every $cT$ according to observer $R_O$'s wristwatch. Then we have a period $cT_e$ of emission by $R_O$, a period $cT_r$ of reception of the light pulse (echo). Under the viewpoint of $R'_O$, at every $cT'$ he (she) receives a pulse of light which is immediately reflected back; so we have, in $WLR'_O$, an "emission" with period $cT'$, according to $R'_O$'s own wristwatch. Following the first postulate of special relativity, the physics under the viewpoint of $WLR'_O$ and $WLR_O$, is the same. Then the ratio $k$ between the periods $cT'$ and $cT_e$ is the same as $cT_r$ and $cT'$.

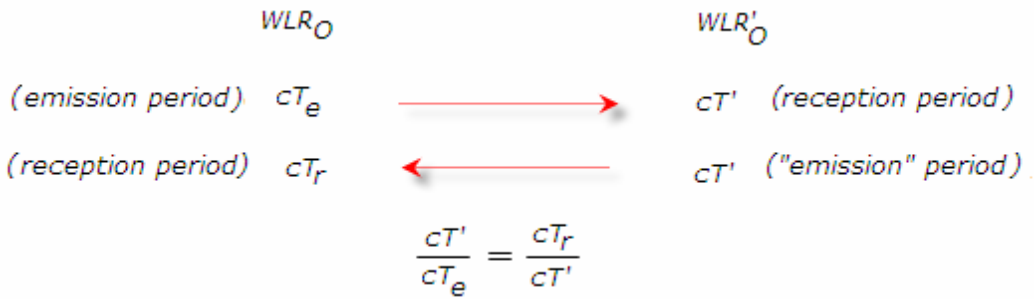

**Figure 4** – The physics is the same for $R_O$ and $R'_O$, then $cT_r/cT'=cT'/cT_e$.

Then we have

$$k = \frac{cT_r}{cT'} = \frac{cT'}{cT_e}, \tag{9}$$

$$\frac{cT_r}{cT'}\frac{cT'}{cT_e} = k^2 = \frac{1+\beta}{1-\beta}, \tag{10}$$

$$k = \sqrt{\frac{1+\beta}{1-\beta}}. \tag{11}$$

This is the so called Doppler factor[2].

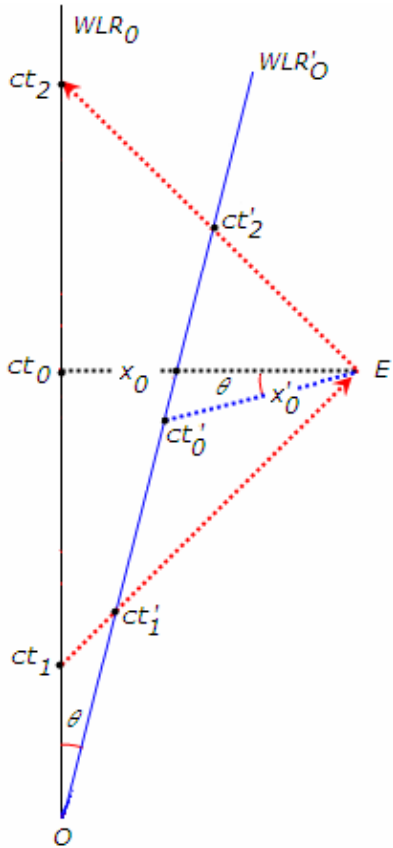 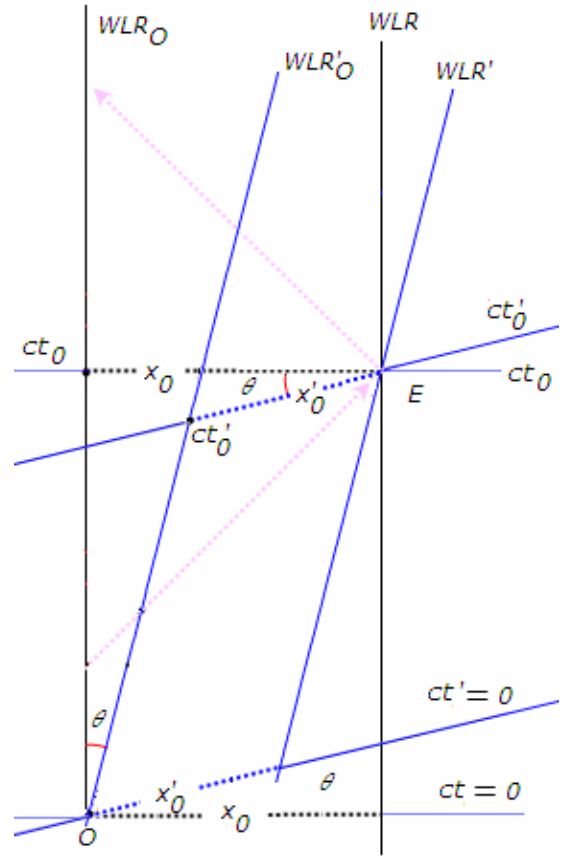

**Figure 5** – Both observers, $R_O$ and $R'_O$, each in his (her) worldline, detect the spacetime coordinates $(x_0, ct_0)$ and $(x'_0, ct'_0)$ of an arbitrary event $E$ through the radar technique.

**Figure 6** - The line containing the segment $x_0$ is a simultaneity line to observers $R_O$ and $R$; all parallel lines to it define different simultaneity lines in the stationary $K$. The line containing the segment $x'_0$ is a simultaneity line for $R'_O$ and $R'$; all parallel lines to it define different simultaneity lines for $K'$.

The Figure 5 shows a situation in which both observers, $R_O$ and $R'_O$, in $WLR_O$ and in $WLR'_O$, respectively, can detect the spacetime coordinates $(x_0, ct_0)$ and $(x'_0, ct'_0)$ of an arbitrary event $E$ using the radar detection procedure. A light pulse which is emitted from the spatial origin of $K$ at time $ct_1$, in $WLR_O$, crosses $WLR'_O$ at time $ct'_1$ and returns to $K$ at $ct_2$ after crossing $WLR'_O$ at time $ct'_2$ in its way back to $WLR_O$. According to (3) and (4), the observer $R_O$ should find $x_0$ and $ct_0$ as

$$x_0 = \frac{1}{2}(ct_2 - ct_1), \qquad (12)$$

$$ct_0 = \frac{1}{2}(ct_2 + ct_1). \qquad (13)$$

According to first postulate of special relativity, $R'_O$ should encounter, with no mystery,

$$x'_0 = \frac{1}{2}(ct'_2 - ct'_1), \qquad (14)$$

$$ct'_0 = \frac{1}{2}(ct'_2 + ct'_1), \qquad (15)$$

where $x'_0$ is the spatial distance between observer $R'_O$ and the event $E$ and $ct'_0$ is the time at which event $E$ occurs, according to $R'_O$. All events occurring on the line which contains the observer $R'_O$ and the event $E$ are said simultaneous to $R'_O$ and $E$. The left side of (14) and (15) are, in fact, the spacetime coordinates, $x'_0$ and $ct'_0$, of event $E$ according to observer $R'_O$.

As we have seen, both observers can measure the spacetime coordinates of an arbitrary event $E$ directly on their own worldlines.

As a consequence of the way in which the moving observer $R'_O$ detects the spacetime coordinates of a given event is that the geometric locus of the events which have, from $R'_O$ point of view, the same space coordinates $x'_0$ is a straight line parallel to his (her) worldline $WLR'_O$ (presented in Figure 6 as $WLR'$) located at a given distance $x'_0$ apart. $WLR'$ is the worldline of an observer $R'$ who is at rest with respect to $R'_O$ in $K'$. Also the geometric locus of the events which have, from $R_O$ point of view, the same space coordinates $x_0$ is a straight line parallel to his (her) worldline $WLR_O$ (shown in Figure 6 as $WLR$). $WLR$ is the worldline of an observer $R$ who is at rest relative to $R_O$ in $K$.

For the same reasons the geometric locus of the events which have the same time coordinate is a straight line that makes an angle $(\pi/2-\theta)$ with the worldline of the stationary observer as shown in Figure 6. In particular, the line which goes through the origin of the space-time diagram represents the geometric locus of the events which have, from the point of view of the moving observer, a zero time coordinate.

## 4. The scale factor

The spacetime diagrams are not Euclidean. In a two-dimensional diagram, the spacetime is fully represented on a Euclidean plane (sheet of paper) because both the Euclidean points and the events can be labeled by pairs of real numbers. However there is an inherent limitation with the use of such diagrams. The Euclidean lengths of lines in the diagrams do not correspond to proper times or proper lengths in spacetime. There is in fact a scale factor which varies according to the direction (velocity) considered.

Let us pick a rule and measure the length of what is the representation of a $WLR'_O$ segment that goes from the origin $O$ up to the point where it is marked $ct'_1$ on Figure 7. Do the same for the segment from $O$ to the point where is marked $ct_1$. Let us name such values as $O1'$ and $O1$, respectively. We are assuming that on $WLR_O$ the scale factor between the true spacetime length $ct_1$ and the corresponding Euclidean length $O1$ is just $1$. Following a simple school geometry rule, we get the ratio between $O1'$ and $O1$,

$$\frac{O1'}{\sin\frac{3\pi}{4}} = \frac{O1}{\sin(\frac{\pi}{4}-\theta)} . \tag{16}$$

$$\frac{O1'}{O1} = \frac{1}{\cos\theta - \sin\theta} \tag{17}$$

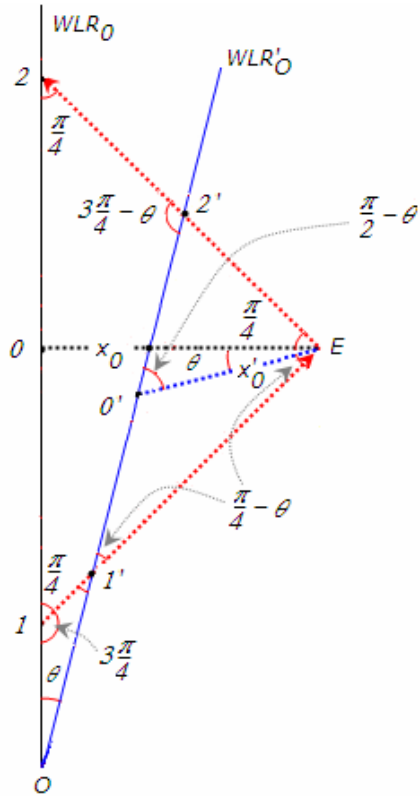

**Figure 7** – The true value of $ct'_1$ is related to the Euclidean length $O1'$ through the scale factor $M$, $ct'_1 = M\,O1'$. By its turn Euclidean length $O1'$ is related to $O1$ through (18).

Since $\theta = arctan\beta$ we get

$$O1' = \frac{\sqrt{1+\beta^2}}{1-\beta} O1 . \tag{18}$$

From (9) (Doppler factor) we can write that the relation between the true spacetime value $ct'_1$, measured by $R'_0$ wristwatch, and $ct_1$, measured by $R_O$ wristwatch, is

$$ct'_1 = ct_1 \sqrt{\frac{1+\beta}{1-\beta}} . \tag{19}$$

Identifying $ct_1$ as being equal to $O1$, by construction, we get from (18) and (19) the following

$$ct'_1 = \frac{1-\beta}{\sqrt{1+\beta^2}} \sqrt{\frac{1+\beta}{1-\beta}} \; O1', \qquad (20)$$

$$ct'_1 = M \; O1', \qquad (21)$$

$$M = \sqrt{\frac{1-\beta^2}{1+\beta^2}} . \qquad (22)$$

$M$ is the scale factor between the true Minkowski spacetime time coordinate value and the Euclidean value length used to represent it on the diagram. Notice that for $\beta=0$, the scale factor reduces to $M=1$ according to our initial assumption above. The space coordinate represented by the Euclidean segment *EO* which is equal to the Euclidean length *02* has obviously the same scale factor *1*.

Now lets us focus our attention on *K'*, the moving inertial reference frame. The event *E* has the spacetime coordinates given by (14) and (15). We have already learned that the scale factor for the time coordinate is *M* given by (22). The *K'* space coordinate $x'_0$ for the same event *E* is represented by the Euclidean segment E0' which is equal to *1'0'=2'0'*; this can be promptly inferred from Figure 7 (angles 1'E0' and E1'0' are both equal to $(\pi/4-\theta)$).

## 5. Length contraction and time dilation

We know that the proper length of a rod is the distance between its ends measured by an observer which is in a state of rest relative to it. In Figure 7 the segment which has length $x_0$ may be "seen" as a rod of proper length $x_0$ in the reference frame *K* where $R_O$ and *R* are at rest. The rod itself lies on a simultaneity line at the viewpoint of *K*. The $WLR_O$ and *WLR* represent the worldlines of both ends of such rod.

In Figure 7 the segment which has length $x'_0$ may be "seen" as a rod of proper length $x'_0$ in the reference frame *K'* where $R'_O$ and *R'* are at rest. The rod itself lies on a simultaneity line at the viewpoint of *K'*. The $WLR_O$ and *WLR* represent the worldlines of the rod both ends.

At this point, we would like to know how a rod at rest in a given reference inertial frame is "seen" by another inertial reference frame moving with constant velocity relative to it.

In Figure 8 we have a rod of proper length $L'_0$ in the moving frame *K'* ($R'_0$ 's inertial frame). To the stationary frame such rod appears to have a length $L_0$. From school geometry, the Euclidean lengths *AE* and *BE* are related by

$$\frac{AE}{\sin(\frac{\pi}{2}-2\theta)} = \frac{BE}{\sin(\frac{\pi}{2}+\theta)} . \qquad (23)$$

$L_0$ is measured in *K* ($R_0$ 's stationary inertial frame); then we can identify *AE* as $L_0$:

$$\frac{L_0}{\cos(2\theta)} = \frac{BE}{\cos(\theta)} , \qquad (24)$$

$$L_0 \frac{1+\beta^2}{1-\beta} = BE\sqrt{1+\beta^2} . \qquad (25)$$

The Euclidean length *BE* is related to *L'₀* through

$$BE = L'_0 \, M \,,. \tag{26}$$

Then we get

$$L_0 \frac{1+\beta^2}{1-\beta} = L'_0 \sqrt{\frac{1+\beta^2}{1-\beta^2}} \sqrt{1+\beta^2} \,, \tag{27}$$

$$L_0 = L'_0 \sqrt{1-\beta^2} \tag{28}$$

So a rod with proper length *L'₀* in *K'*, the moving inertial reference frame, seems to have length equal to $L_0 < L'_0$ to observer in *K*. This is the length contraction relativistic effect.

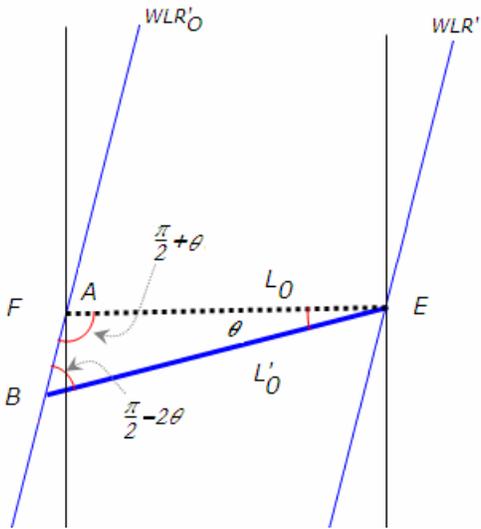
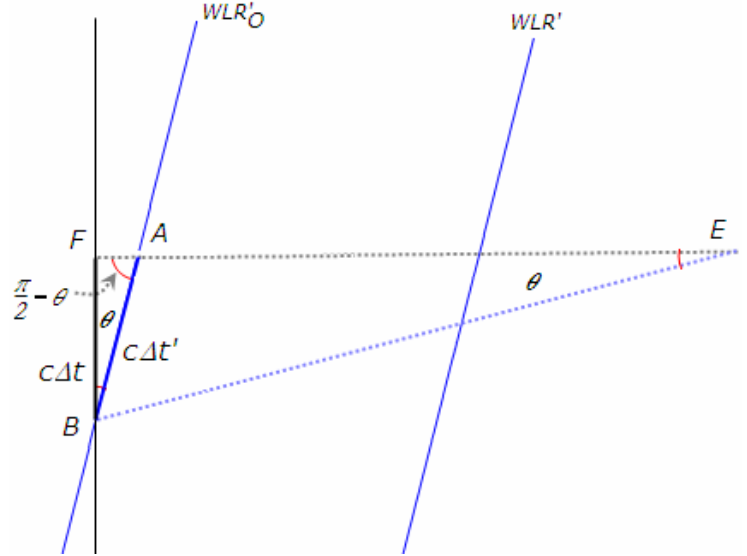

**Figure 8** - The Euclidean line segments *BE* and *AE* represent, respectively, the proper length of a rod at rest in *K'* and length "seen" by observer in *K*.

**Figure 9** - The Euclidean line segment *AB* represents the time interval *cΔt'* as measured by $R'_0$, in *K'* while the Euclidean line segment *FB* represents the time interval *cΔt* as measured by $R_O$, in *K*.

Consider now the Figure 9. On the worldline *WLR'ₒ* we pick the Euclidean line segment *AB* which represents the time interval *cΔt'*, as measured by observer $R'_0$, in *K'*. The Euclidean line segment *FB* represents the time interval *cΔt* as "seen" in *K*. The points *F* and *A* belong to a simultaneity line in *K*. The trivial school geometry says that

$$FB = AB \cos\theta \tag{29}$$

where *cosθ* is expressed in terms of *β*

$$\cos\theta = \frac{1}{\sqrt{1+\beta^2}} \,. \tag{30}$$

Since the scale factor for *FB* is *M=1*, we write

$$FB = c\Delta t \tag{31}$$

The scale factor for *AB* is given by (22), then

$$AB = \frac{c\Delta t'}{M}. \tag{32}$$

Combining (29) (31) and (32) we get

$$c\Delta t = c\Delta t' \frac{1}{\sqrt{1-\beta^2}}. \tag{33}$$

This is the time dilation effect. The elapsed time *c∆t'* in *K'* is "seen" by *K* as *c∆t>c∆t'*.

## 6. Lorentz Transformations

With the scale factor *M* in mind it is possible to derive also the Lorentz transformation[2] in terms of the actual Euclidean lengths on the drawing of the diagram. So from the Figure 7, instead of (14) and (15), we would have

$$E0' = \frac{1}{2}(O2' - O1') \tag{34}$$

$$OO' = \frac{1}{2}(O2' + O1'). \tag{35}$$

According to (18) we then would write

$$ct_1 = O1' \frac{1-\beta}{\sqrt{1+\beta^2}}. \tag{36}$$

From Figure 7, we have

$$ct_2 = O2'(\cos\theta + \sin\theta) = O2' \frac{1+\beta}{\sqrt{1+\beta^2}}. \tag{37}$$

From the above, (12) and (13) may be put in the following way

$$x_0 = \frac{1}{2}\frac{1}{\sqrt{1+\beta^2}}((O2' - O1') + \beta(O2' + O1')) \tag{38}$$

$$ct_0 = \frac{1}{2}\frac{1}{\sqrt{1+\beta^2}}((O2' + O1') + \beta(O2' - O1')) \tag{39}$$

or

$$x_0 = \frac{1}{2}\frac{1}{\sqrt{1+\beta^2}}(E0' + \beta OO') \tag{40}$$

$$ct_0 = \frac{1}{2}\frac{1}{\sqrt{1+\beta^2}}(O0' + \beta E0') \tag{41}$$

Taking account the scale factor M given by (24)

$$O0' = M\, ct'_0 \tag{42}$$

$$E0' = M\, x'_0 \tag{43}$$

and the true Lorentz Transformations for the event E are

$$x_0 = \frac{1}{\sqrt{1-\beta^2}}(x'_0 + \beta\, ct'_0) \tag{44}$$

$$ct_0 = \frac{1}{\sqrt{1-\beta^2}}(ct'_0 + \beta\, x'_0) \tag{45}$$

Since E is an arbitrary event, we drop the subscript

$$x = \frac{1}{\sqrt{1-\beta^2}}(x' + \beta\, ct') \tag{46}$$

and

$$ct = \frac{1}{\sqrt{1-\beta^2}}(ct' + \beta\, x'), \tag{47}$$

recovering the Lorentz-Einstein transformations for the space time coordinates of the same event detected by observers of the two involved inertial reference frames.

## 7. Conclusions

Starting with the Euclidean representation of the relativistic spacetime diagram for the two-dimensional case, we show how a student can have a better insight of the relative "size" of proper times and proper lengths in special relativity. He (she) can measure the lengths of line segments on the diagram plot and through the use a scale factor to have a better idea of the relative magnitudes of the spacetime coordinates:

$$(\text{true physical magnitude}) = \left(\sqrt{\frac{1-\beta^2}{1+\beta^2}}\right)(\text{read from diagram magnitude})$$

We show that length contraction, time dilation and the Lorentz Transformations formulas can be obtained with the help of simple trigonometry properties.

---

[1] Nilton Penha and Bernhard Rothenstein, "*On spacetime coordinates in special relativity*", http://arxiv.org/ftp/arxiv/papers/0705/0705.0941.pdf.
[2] R. d'Inverno, *Introducing Einstein's Relativity,* (Oxford University Press 1992) p20 and 29.